\documentclass{aa}
\usepackage{epsfig}
\usepackage{natbib}
\bibpunct{(}{)}{;}{}{,}
\newif\ifAMStwofonts
\def\spose#1{\hbox to 0pt{#1\hss}}
\def\simlt{\mathrel{\spose{\lower 3pt\hbox{$\mathchar"218$}}
     \raise 2.0pt\hbox{$\mathchar"13C$}}}
\def\simgt{\mathrel{\spose{\lower 3pt\hbox{$\mathchar"218$}}
     \raise 2.0pt\hbox{$\mathchar"13E$}}}

\begin{document}
%\thesaurus{02.08.1; 09.01.1; 11.05.2}
\title{Self-enrichment in globular clusters. I. An analytic approach}
\author{S. Recchi\inst{1,2,3}\thanks{recchi@astro.univie.ac.at} 
        \and I.J. Danziger\inst{3}\thanks{danziger@ts.astro.it} 
        }
\offprints{S. Recchi}
\institute{
University Observatory of Vienna, T\"urkenschanzstrasse 17, A-1180 
Vienna, Austria\and
Institut f\"ur Theoretische Physik und Astrophysik, Kiel University, 
Olshausenstrasse 40, 24098 Kiel, Germany\and
INAF, Dipartimento di Astronomia, Universit\`a di Trieste, Via G.B. 
Tiepolo, 11, 34131 Trieste, Italy}
\date{Received  /  Accepted   }

\abstract{ By means of analytical calculations, we explore the
self-enrichment scenario for Globular Cluster formation.  According to
this scenario, an initial burst of star formation occurs inside the
core radius of the initial gaseous distribution.  The
outward-propagating shock wave sweeps up a shell in which
gravitational instabilities may arise, leading to the formation of a
second, metal-enriched, population of stars.  We find a minimum mass
of the proto-globular cluster of the order of 10$^6$ M$_\odot$.  We
also find that the observed spread in the Magnitude-Metallicity
relation can be explained assuming cluster-to-cluster variations of
some parameters like the thermalization efficiency, the mixing
efficiency and the Initial Mass Function, as well as variations of the
external pressure.   } \maketitle

\bigskip\bigskip

\section{Introduction}

Globular Clusters (GCs) are among the oldest objects in the Galaxy and
thus are fossil records of its formation.  Their age is comparable
with the age of halo stars but, quite surprisingly, the lowest
metallicity found in GCs ([Fe/H] $\simlt$ $-2.5$) is more than two
orders of magnitude larger than the iron abundance of the most
metal-poor halo stars (e.g. the recently discovered star HE 0107--5240
has a metallicity [Fe/H] $=-5.3$; Christlieb et al. 2004).  This could
be only a statistical effect, since GCs make up only $\sim$ 1\% of the
stars in the halo of the Galaxy.  Moreover, the metallicity
distribution of the halo stars and of the GCs peak at the same value
(Ashman \& Zepf 1998).  This could mean that the halo stars are the
debris of the tidal disruption of cluster-like objects.  Observations
of the distribution of a wide sample of stars detected in the Sloan
Digital Ski Survey (Newberg et al. 2002) seems to support this
picture.
  
More detailed analyses of the metallicity distribution of halo stars
and GCs (Laird et al. 1993; Carney et al. 1996) indicate that there is
a statistically significant difference between the two metallicity
distributions that cannot be explained by the lower numbers of GCs.
Moreover, field halo stars are almost all CN-weak, whereas some GCs
show a bimodal CN distribution, with some stars having high CN
abundances (Norris \& Smith 1981).  This may indicate a difference in
the formation process of these two classes of objects.

In many GCs the abundances of elements from C to Al show complex
patterns and star-to-star abundance variations, such as the well-known
O-Na and Mg-Al anticorrelations (Gratton et al. 2001; Gonzalez \&
Wallerstein 1998).  A few GCs: $\omega$ Centauri (Freeman \& Rodgers
1975; Smith et al.  2000; Pancino et al.  2002), M22 (Norris \&
Freeman 1983; Lehnert, Bell \& Cohen 1991) and maybe M92 (Langer et
al. 1998) show a significant spread even in iron-peak abundances, with
signatures of a metallicity gradient.  Pancino et al. (2002) found
also in $\omega$ Centauri that [Fe/H] correlates with [Cu/Fe], whereas
the [Si/Fe] vs.  [Fe/H] plot shows a plateau until [Fe/H] $\sim$ $-1$
and then an anticorrelation.  These patterns are characteristic of
iron enrichment from SNeIa (Matteucci et al.  1993).

The range of metallicities in GCs is similar to that found in Local
Group Dwarf Spheroidals (DSphs) and even in this case no galaxies with
[Fe/H] $\simlt$ $-2.2$ are found.  This may indicate a common origin
of these two classes of objects, or at least a similar enrichment
scenario.

In order to explain the lack of metal-poor GCs, two processes have
been invoked: pre-enrichment from PopIII stars (Beasley et al. 2003)
and the {\it self-enrichment} scenario, the second being the only one
able to explain abundance variations observed in some GCs.  The idea
behind this is that, if a proto-globular cluster (PGC) is massive
enough, it may retain the heavy elements produced by the first
supernovae exploding after the collapse of the cloud.  These metals
may mix with the surrounding, unpolluted medium and any following
episodes of star formation arise from a metal-enriched gas.  The
pioneers in this kind of study have been Dopita \& Smith (1986) and
Cayrel (1986), but many other authors have studied the subject (Morgan
\& Lake 1989; Ikuta \& Arimoto 2000; Smith 2000 among others).  In
particular, Shustov \& Wiebe (2000; hereafter SW00) and Parmentier et
al. (1999; hereafter P99) addressed the question of how massive a PGC
should be in order to retain the products of at least one supernova
(SN) and how many SNe can a PGC sustain before being disrupted as a
result of the injected energy.  P99 tried also to compare the results
of their model with the magnitude-metallicity relation in a particular
sample of Milky Way GCs (namely the older ones).

This is the first of a series of three papers dealing with analytical
and numerical calculation of the self-enrichment process in GCs.  In
this paper, after carefully studying the available data about GCs and
Local Group DSphs (Section 2), we perform an analytical study of the
evolution of a PGC, taking into consideration the possibility of
triggered star formation (Section 3).  Results are presented and
discussed in Section 4.  Finally, we draw our conclusions in Section
5.  In the forthcoming papers, we will analyze 1-D (paper II) and 2-D
(paper III) chemodynamical simulations of the evolution of a PGC.

\section{On the mass dependence in the metallicity distribution of 
Milky Way GCs}

The most obvious reference for any study of Milky Way GCs is the
Harris Catalogue (http://physun.physics.mcmaster.ca/Globular.html; see
also Harris 1996).  The metallicity-luminosity diagram for the whole
galactic globular cluster system is shown in Fig. 1.  We include also
in the same plot the metallicity-luminosity relation for Local Group
DSphs, taken from the Mateo (1998) review.

\begin{figure}
\centering
\vspace{-1cm}
\epsfig{file=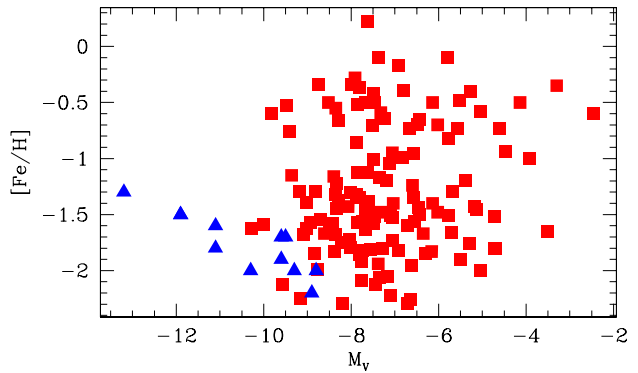, height=9cm,width=9cm}
\vspace{-1cm}
\caption[]{\label{fig:fig 1} Plot of the total visual magnitude M$_V$
vs.  metallicity [Fe/H] for GCs of the Milky Way (squares) and Local
Group DSphs (triangles).  Data are from the Harris Catalogue (GCs) and
from Mateo (1998) (DSphs).}
\end{figure}

This plot does not show any obvious correlation among GCs, whereas the
DSph system exhibits a well defined and well known trend, such that
the faintest DSphs are also the most metal-poor.  The classical
explanation of this correlation is that the effect of galactic winds
decreases with the increasing depth of the galactic potential well
(Larson 1974; Dekel \& Silk 1986).  This mechanism cannot easily be
extended to GCs, mostly because the DSphs show a very large
mass-to-light ratio, presumably a hint of dominant dark matter halos,
whereas GCs show very little evidence of dark matter.

There have been several attempts in the literature to calibrate the
integrated light of GCs in order to recover the metallicity (Danziger
1973; Zinn 1980).  The Harris catalogue is based on the Zinn \& West
(1984) calibration, but this is not the only one and care is needed
before drawing conclusions from the inspection of the
metallicity-luminosity plot.  In particular, Carretta \& Gratton
(1997) proposed a revised version of the Zinn \& West (1984)
calibration.  This new calibration changes the resulting metallicity
in particular in the range of high abundances.  The
metallicity-luminosity relation with this calibration is shown in
Fig. 2.

\begin{figure}
\centering
\vspace{-1cm}
\epsfig{file=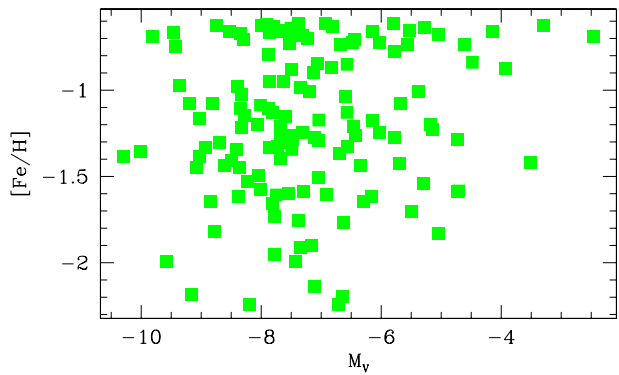, height=9cm,width=9cm}
\vspace{-1cm}
\caption[]{\label{fig:fig 2} Plot of the total visual magnitude M$_V$ vs. 
metallicity [Fe/H] for GCs of the Milky Way, by means of the calibration 
proposed by Carretta \& Gratton (1997).}
\end{figure}

It has been suggested that the mass loss due to stripping with passage
through the disk of the Galaxy might dominate the mass budget of GCs.
This could be the reason why there is such an enormous spread in this
relation.  One may suppose that the galaxies with large $|b|$, thus at
large distance from the disk of the Galaxy, may have suffered less
stripping.  For this reason, we select the GCs with $|b| > 20$ and
plot in Fig. 3 the metallicity-luminosity relation for this selected
sample of GCs, with both the Zinn \& West (1984) calibration (filled
squares) and the Carretta \& Gratton (1997) one (open squares).

\begin{figure}
\centering
\vspace{-1cm}
\epsfig{file=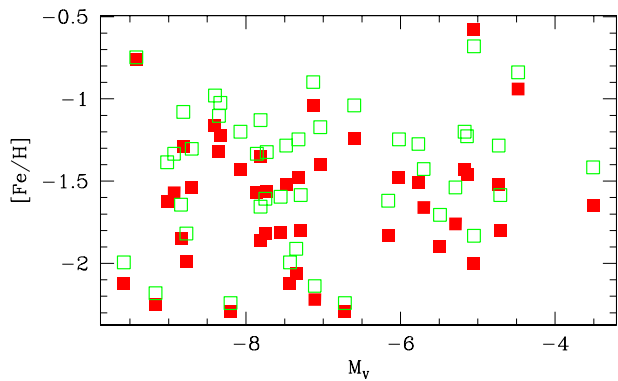, height=9cm,width=9cm}
\vspace{-1cm}
\caption[]{\label{fig:fig 3} Plot of the total visual magnitude M$_V$
vs.  metallicity [Fe/H] for the GCs of the Milky Way with $|b| > 20$.
The filled squares are relative to the Zinn \& West (1984)
calibration, the open squares to the Carretta \& Gratton (1997) one.}
\end{figure}

Although Fig. 3 shows a reduced spread compared to Figs. 1 and 2, an
obvious correlation still cannot be detected, suggesting that, besides
external effects, internal properties of the GCs should also be
invoked in order to justify this enormous spread.

For a reduced sample of GCs, mass measurements are available.  In this
case, we do not have the problem to calibrate the metallicity index,
but the determination of the mass of GCs is very uncertain and
model-dependent (Pryor \& Meylan 1993).  We adopt the same selection
criterion introduced for Fig. 3 (i.e. selection of GCs at large
galactic latitudes) and we plot in Fig. 4 the GC masses as a function
of the metallicity for the GCs studied by Pryor \& Meylan (1993).
Also in this case, no obvious trend can be identified, unless we
exclude from the plot the two most massive GCs.  In this case we can
see that the metallicity slightly increases with mass.

\begin{figure}
\centering
\vspace{-1cm}
\epsfig{file=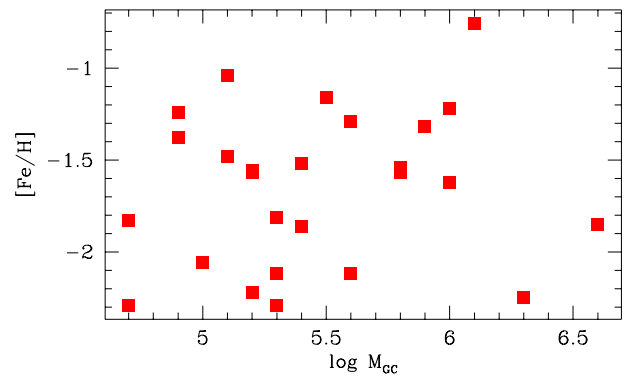, height=9cm,width=9cm}
\vspace{-1cm}
\caption[]{\label{fig:fig 4} Plot of the total GC mass log M$_{\rm GC}$ 
vs.  metallicity [Fe/H] for the GCs of the Milky Way with $|b| > 20$.  
Data are taken from Pryor \& Meylan (1993).}
\end{figure}

\section{An analytic approach}

\subsection{Set-up of the model}

Let us assume an isothermal sphere ($P\sim \rho$) in hydrostatic
equilibrium.  The hydrostatic equilibrium, in spherical symmetry, is
expressed by the formula ${dP\over dr}=-{{G M(r) \rho(r)}\over r^2}$,
whereas the conservation of mass implies that ${dM \over dr} = 4 \pi
\rho (r) r^2$.  Combining these two equations, we obtain the
fundamental equation of equilibrium:

\begin{equation}
{1\over r^2}{d\over dr}\biggl( {r^2 \over \rho} {dP \over dr}\biggr) 
=-4 \pi G \rho(r).
\end{equation}
\noindent
We can modify this equation, introducing two dimensionless variables

\begin{equation}
\cases{\omega = -\ln {\rho/\rho_{\rm 0}}\cr
\xi = A r, \;\; A^2={{4 \pi G \rho_{\rm 0}}\over {c_s}^2},\cr}
\end{equation}
\noindent
where $c_s$ is the sound speed, obtaining:

\begin{equation}
{d^2\omega \over {d \xi^2}}+{2\over \xi}{d\omega \over d\xi}=e^{-\omega},
\end{equation}
\noindent
This is the classical Lane-Emden equation for an isothermal sphere and
cannot be solved analytically.  With a simple, second-order numerical
integration, we obtain the density profile plotted in Fig. 5.  Also
plotted (dotted line) is the asymptotic trend at large $\xi$ ($\rho
\sim \rho_0 \cdot \xi^{-2}$)

\begin{figure}
\centering
\vspace{-0.5cm}
\epsfig{file=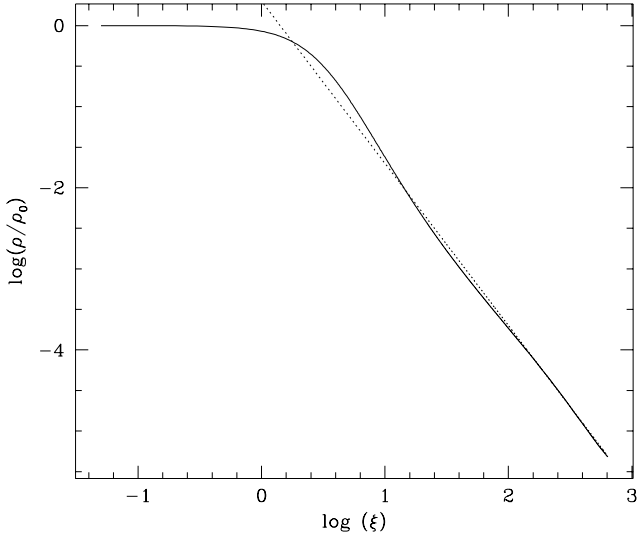, height=9cm,width=9cm}
\vspace{-0.5cm}
\caption[]{\label{fig:fig 5} Normalized density profile in an isothermal 
sphere.  Solid line: solution of the numerical integration; dotted line: 
asymptotic trend at large $\xi$.}
\end{figure}

This density profile can be well approximated by a King profile:

\begin{equation}
\rho(r) = \cases{\rho_0  \;\;\;\;\;\;\; {\rm if} \;\; r < r_c \cr
\rho_0 {{{r_c}^2} \over {r^2}}  \;\; {\rm if} \;\; r \geq r_c,\cr}
\end{equation}
\noindent
where $r_c$ is the core radius and the profile is truncated at a tidal
radius $R$.  The concentration parameter $c = \log (R / r_c)$ is
$\sim$ 1 for most of metal-poor GCs (see Harris Catalogue), thus we
can consider hereinafter $r_c = 0.1 \, R$.  We assume pressure
equilibrium between the PGC and the external medium, thus, defining
$P_h$ as the external pressure, we obtain $\rho(R) = 0.01 \, \rho_0 =
P_h / {c_s}^2$.  We use this expression to calculate the mass of the
PGC as a function of $R$ and $P_h$.  Finally, this expression can be
used to obtain the tidal radius $R$ as a function of the total PGC
mass and the external pressure.  The final expression is:

\begin{equation}
R = \Biggl\lbrack {1 \over 2.8} {{3 k T} \over {4 \pi \mu m_H}} 
\Biggr\rbrack^{1/3} \Biggl({M \over P_h}\Biggr)^{1/3}.
\end{equation}
\noindent
Assuming $\mu = 1.3$ and $T = 10^4$ (as we will do hereinafter), we 
obtain:

\begin{equation}
R_{100} = 1.385 \Biggl({M_6 \over P_4}\Biggr)^{1/3},
\end{equation}
\noindent
where $R_{100}$ is in units of 100 pc, $M_6$ in units of 10$^6$ M$_\odot$ 
and $P_4$ in units of 10$^4 k$.  The dependence of $\rho_0$ with $P_h$ is 
$\rho_{0,22} = 2.17 P_4$, where $\rho_{0,22}$ is in units of 10$^{-22}$ 
g cm$^{-3}$.

\subsection{The collapse of the PGC}

Let us now assume that the sphere loses its equilibrium and starts
collapsing.  This assumption is justified by the fact that in the
eq. (1) we neglected the cooling, thus the configuration should lose
the equilibrium due to radiative losses.  The cooling time in the
inner core ($t_c \simeq 3 k T / (2 n \Lambda (T))$, where $\Lambda
(T)$ is the cooling function) is approximatively 2 $\cdot 10^4$ yr
with our assumptions.  The collapse of the core will be much faster
than the collapse of the external mantel.  If we assume for simplicity
the free fall time-scale of the mantel as $t_{ff, man} = \sqrt {[3 \pi
/(32 G \overline \rho_{man})]}$, where $\overline \rho_{man} = 3
\rho_0 r_c^2 (R - r_c) / (R^3 - r_c^3)$ is the average density of the
external mantel, we obtain $t_{ff, c} / t_{ff, man} \sim 0.16$;
that is, if we assume for simplicity two separate collapse episodes, the
mantel will not have time to loose its equilibrium configuration in
the interval of time in which the core collapses.  We can thus assume
that a fraction of the core collapses and forms stars in an
instantaneous burst of star formation, while the external mantel keeps
its starting density profile.

The efficiency of star formation in the core (namely the fraction of
core mass turned into stars) is assumed to be 0.3, consistent with the
star formation efficiencies obtained in the cores of dense clouds by
Bate, Bonnell \& Bromm (2003).

The mass of stars in this first, almost metal-free, episode of star
formation is thus:

\begin{equation}
M_{*, 6} = 1.071 \cdot 10^{-2} M_6.
\end{equation}
\noindent
The number of SNeII is given by: $N_{SNII} = \int_8^{100} A m^{-x}
dm$, where $A$ is calculated from $M_* = \int_{m_l}^{100} A m^{1 - x}
dm$.  A Salpeter ($x = 2.35$) index can be taken as a good
approximation to the IMF slope in clusters (Kroupa 2002).  $m_l$ is
the lower cut-off of the IMF.  We define $g(m_l) = m_l^{-0.35} -
100^{-0.35}$ and we take $m_l$ as a free parameter.  We obtain:

\begin{equation}
N_{SNII} = 162.1 {M_6 \over g(m_l)}.
\end{equation}
\noindent
For a reasonable choice of $m_l$ (i.e. $m_l = 1$) this gives
$N_{SNII}$ $\simeq$ 200 $M_6$, thought by P99 to be the critical
number of SNe sustainable against PGC disruption.  Assuming the rate
of SNeII constant in time, we can calculate the luminosity of the
burst.  This is given by: $L = \eta E_{SNII} N_{SNII} / \Delta t$,
where $\eta$ is the fraction of the mechanical energy of the explosion
which goes into thermal energy of the medium (the so-called {\it
thermalization efficiency}), $E_{SNII}$ is the energy of a single SNII
(assumed to be 10$^{51}$ erg) and $\Delta t \sim 30$ Myr is the
interval of time in which SNeII explode.  The thermalization
efficiency can be evaluated assuming the evolution of a SNR in a
uniform medium.  We can calculate the time at which the expansion
velocity of the SNR becomes equal to the local sound speed, the
kinetic energy of the shell at this time, and divide this energy by
the blast wave energy (see Cioffi, McKee \& Bertschinger 1988;
Bradamante, Matteucci \& D'Ercole 1998; Recchi, Matteucci \& D'Ercole
2001 for more details).  We obtain:

\begin{equation}
\eta \simeq 0.02 \; n_0^{-54/49} \xi^{-15/98} c_{0,6}^{5/7},
\end{equation}
\noindent
where $\xi$ is the metallicity (in units of Z$_\odot$) and $c_{0,6}$
is the local sound speed (in units of 10$^6$ cm s$^{-1}$).  Assuming
$\xi = 10^{-4}$ (typical of the most metal-poor halo stars), we
obtain:

\begin{equation}
\eta = 4.36 \cdot 10^{-4} P_4^{-54/49}.
\end{equation}
\noindent
It is worth mentioning that the determination of $\eta$ in this way is
too simplistic and many other physical processes contribute to its
determination (see Melioli \& de Gouveia Dal Pino 2004).  Therefore we
will keep $\eta$ in this paper as a free parameter, bearing in mind
that, at least in the initial stages of the evolution of the system,
its value might be very low.  The luminosity of the burst (in units of
10$^{38}$ erg s$^{-1}$) is thus given by:

\begin{equation}
L_{38} = 1.712 M_6 \Biggl({\eta \over g(m_l)}\Biggr).
\end{equation}
\noindent
We assume that mass and energy of this first generation of stars are
released in a spherical volume of radius $r_{SF} << R$ and we keep the
$r^{-2}$ gas density profile outside it.

\begin{table}
\begin{centering}
\caption[]{Ejecta fraction and mass in metals}
\begin{tabular}{ccc}
\noalign{\smallskip}
\hline
\noalign{\smallskip}
M$_{in}$ & $f_{ej}$ & $m_z$\\
\noalign{\smallskip}
\hline\noalign{\smallskip}
12 & 0.90 & 0.48\\
13 & 0.89 & 0.50\\
15 & 0.90 & 0.73\\
18 & 0.92 & 1.58\\
20 & 0.90 & 2.34\\
22 & 0.91 & 2.74\\
25 & 0.93 & 4.06\\
30 & 0.94 & 6.01\\
35 & 0.92 & 7.18\\
40 & 0.90 & 8.22\\
\noalign{\smallskip}
\hline
\end{tabular}
\vspace{0.3cm}

Total ejected fraction ($f_{ej}$ = 1 - M$_{rem}$/M$_{in}$) and mass of 
metals $m_z$ as a function of initial mass.  Data from Woosley \& 
Weaver 1995 (Z=10$^{-4}$ Z$_{\odot}$; case B)
\end{centering}
\end{table} 

\subsection{Slow winds or fast winds?}

The evolution of wind-blown bubbles and superbubbles has been
carefully analyzed by Koo \& McKee (1992a; 1992b).  Since the results
of these authors will be extensively used in the paper, we recall here
briefly a few results about bubble expansion.  The freely expanding
wind produced by the starburst interacts supersonically with the
unperturbed ISM, creating a classical bubble structure (Weaver et al.
1977), in which two shocks are present.  The external one (``ambient
shock'') propagates through the ISM and creates an expanding cold and
dense shell.  A second shock (``wind shock'') propagates inwards,
thermalizing the impinging wind and creating the hot, rarefied gas of
the bubble interior.  The shocked ambient medium and the shocked wind
are separated by a contact discontinuity.  Initially, the wind density
is large and the velocity of the wind shock is small, therefore the
wind shock is radiative.  At later stages, the wind density decreases
and the wind shock velocity increases, therefore the wind shock is
expected to become adiabatic.  If the energy input is powerful enough,
this transition could occur in an early stage of the bubble expansion,
when the wind is still freely expanding.  In this case, the radiative
cooling of the cavity is dynamically unimportant.  Koo \& McKee
(1992a) call these kinds of winds {\it fast winds}.  The {\it slow
  winds} are instead the ones in which the transition from a radiative
to an adiabatic wind shock occurs later, making the cooling of the
bubble dynamically relevant.
  
Considering a generic power-law density distribution of the ambient
medium $\rho (r) = \rho_{01} r^{-k_\rho}$ and a generic rate of energy
injection $L_{in} = {\cal L} t^{\eta_{in} - 1}$, Koo \& McKee (1992b)
were able to calculate a {\it critical velocity}, namely the velocity
which divides the two types of winds.  This expression is given by:

\begin{equation}
v_{cr} = \biggl\lbrack \biggl( {{\cal L} \over {2 \pi \eta_{in}}} \biggr)
^{1 - k_\rho} \biggl({3 \rho_{01} \over {3 - k_\rho}}\biggr)^{2 - \eta_{in}} 
{1 \over C_1^{3 - k_\rho - \eta_{in}}} \biggr\rbrack
^{\alpha_{k, \eta}},
\end{equation}
\noindent
where $\alpha_{k, \eta} = 1/[14 - 6 k_\rho - (3 + k_\rho) \eta_{in}]$
and $C_1 = 6 \times 10^{-35}$ g cm$^{-6}$ s$^4$.

Assuming a $r^{-2}$ profile and a constant injection rate (i.e.
$\eta_{in} = 1$) eq. (12) becomes:

\begin{equation}
v_{cr} = \Biggl({{\cal L} \over 6 \pi \rho_0 r_c^2} \Biggr)^{1/3}.
\end{equation}
\noindent
By substituting our parameters, we obtain:

\begin{equation}
v_{cr, 8} = 2.84 \cdot 10^{-2} \, \Biggl({M_6 \over P_4}\Biggr)^{1/9} 
\Biggl({\eta \over g(m_l)}\Biggr)^{1/3}.
\end{equation}
\noindent
An approximate value of the velocity of the ejecta from SNeII is given
by $v_{ej} = \sqrt {2 E_{SNII} / M_{ej,SNII}}$, where $E_{SNII} = \eta
\cdot N_{SNII} \cdot 10^{51}$ erg is the total energy released by SNeII,
whereas $M_{ej,SNII}$ is the total mass restored by these SNe.  This
mass is a fraction $f_{ej}$ of the total mass in stars in the interval
[8, 100] M$_\odot$.  Assuming a Salpeter IMF, the mass of stars in
this interval is 0.283 / $g(m_l)$ times the total mass of stars,
whereas $f_{ej}$ can be evaluated from the tables of Woosley \& Weaver
(1995).  Assuming an abundance of the PGC of 10$^{-4}$ Z$_\odot$,
$f_{ej} \sim 0.9$, with almost no dependence on the initial mass (see
Table 1), we thus obtain:

\begin{equation}
v_{ej, 8} = 2.44 \sqrt \eta,
\end{equation}
\noindent
where $v_{ej, 8}$ is in units of 10$^8$ cm s$^{-1}$.  

The ratio $v_{ej} / v_{cr}$ is given by:

\begin{equation}
{v_{ej} \over v_{cr}} = 85.9 \Biggl({M_6 \over P_4}\Biggr)^{-1/9} 
\eta^{1/6} g(m_l)^{1/3},
\end{equation}
which is very large and varies very little with the parameter space.
Even assuming the (very low) thermalization efficiency found in eq.
(10), assuming $m_l = 1$ and neglecting the very weak dependence on
$M_6$ and $P_4$, we obtain $v_{ej} / v_{cr} \simeq 20$.  We can thus
safely state that the winds occurring in PGC are {\it fast winds}.

\subsection{The evolution of the Superbubble}

Koo \& McKee (1992a) defined the {\it fiducial radius} $R_f$ as the
radius at which the wind density equals the ambient density.  For a
general power-law density distribution and a general rate of energy
injection, they obtained:

\begin{equation}
R_f = \biggl\lbrack {{(3 - k_\rho) {\cal L}} \over {6 \pi \eta_{in} 
\rho_{01} v_{in}^{2 + \eta_{in}}}} \biggr\rbrack^{1 \over {3 - k_\rho 
- \eta_{in}}}.
\end{equation}
\noindent
Koo \& McKee (1992) were able to derive the expansion law of the
bubble under various initial conditions and in different stages of the
bubble evolution, as a function of $R_f$.  Indeed, ours is a special
case, since, for $k_\rho = 2$ and $\eta_{in} = 1$ this normalization
breaks down and the fiducial radius approaches zero.  This special
case is called {\it constant velocity bubble} and, physically, this
breakdown corresponds to the fact that the ratio between swept-up mass
and ejected mass is constant.

The ambient medium always dominates these kinds of winds and the
bubble always stays radiative.  The superbubble expands with a
constant velocity $v_b = v_{ej} I / (1 + I)$, where $I = \sqrt {{\cal
    L} / (2 \pi \rho_0 r_c^2 v_{ej}^3)} = \sqrt 3 \cdot (v_{cr} /
v_{ej})^{3/2}$.  With our parameters:

\begin{equation}
v_{b, 6} \simeq 0.532 \Biggl({M_6 \over P_4}\Biggr)^{1/6} \eta^{1/4} 
g(m_l)^{-1/2},
\end{equation}
\noindent
(where $v_{b, 6}$ is in units of 10$^6$ cm s$^{-1}$) and consequently 
the superbubble expands with a law:

\begin{equation}
R_{s, 100} = 5.44 \cdot 10^{-2} \, \Biggl({M_6 \over P_4}\Biggr)^{1/6} 
\eta^{1/4} g(m_l)^{-1/2} \, t_6,
\end{equation}
\noindent
with $t_6$ in units of 10$^6$ yr.  Since the burst originates in a
spherical region of radius $r_c$, $R_{s}$ should be considered as the
difference between the shock radius and $r_c$.  Note that, if the
bubble evolves into a partially radiative bubble (i.e. a bubble in
which most of the shocked wind has cooled down, but most of the bubble
volume is filled with the most recently shocked and still hot portion
of the wind), the velocity would be no more constant, but increasing
with $t^{1/3}$.  However, such an accelerating bubble would produce
Rayleigh-Taylor instabilities with the overlying gas, resulting in a
larger cooling of the shocked wind, preventing the onset of the
partially radiative bubble (see Koo \& McKee 1992b).  Therefore, the
assumption of a constant-velocity bubble is reliable.  The time needed
to reach the tidal radius $R$ is:

\begin{equation}
t_{tid, 6} = 22.9 \Biggl({M_6 \over P_4}\Biggr)^{1/6} 
\eta^{-1/4} g(m_l)^{1/2},
\end{equation}
\noindent
which is of the order of the lifetime of a 8 M$_\odot$ star.  It
implies that the assumption of a constant luminosity is reliable.

The swept-up mass as a function of time is given by:

\begin{equation}
M_{sw, 6} = 4.2 \cdot 10^{-2} M_6^{5/6} P_4^{1/6} \eta^{1/4} 
g(m_l)^{-1/2} t_6.
\end{equation}
\noindent
It is easy to check that the swept-up mass at $t_{tid}$ is similar to 
the initial mass; the difference being the mass in stars.

The mass $m_Z$ of metals ejected in the ISM by a SN whose progenitor
mass is $M_{in}$ is shown in Table 1.  This dependence can be
approximated by:

\begin{equation}
m_Z (M_{in}) = 0.015 (M_{in} - 8)^2 - {2 \over 3} \cdot 10^{-5} 
(M_{in} - 8)^4,
\end{equation}
(see Fig. 6).  This approximation is valid until $M_{in} = 40$
M$_\odot$.  We will assume a constant $m_Z$ above 40 M$_\odot$.  
The total mass of metals released into the ISM as a function of time 
can be expressed as:

\begin{eqnarray}
\lefteqn \,\, M_Z (t) & = & {0.35 \over g(m_l)} M_* \int_{100}^{m(t)} 
M_{in}^{-2.35} m_Z (M_{in}) d M_{in} \nonumber \\ 
\,\, & = & {0.35 \over g(m_l)} M_* \, \Delta_Z (t).
\end{eqnarray}

\noindent
This integral cannot be evaluated analytically.  We solve it
numerically.  We assume a simple law for the stellar lifetimes: $t(m)
= 1200 m^{-1.85} - 3$ Myr (Padovani \& Matteucci 1993).  The resulting
$\Delta_Z (t)$ is shown in Fig. 6 and the values every Myr are
tabulated in Table 2.

\begin{figure}
\centering
\vspace{-1cm}
\epsfig{file=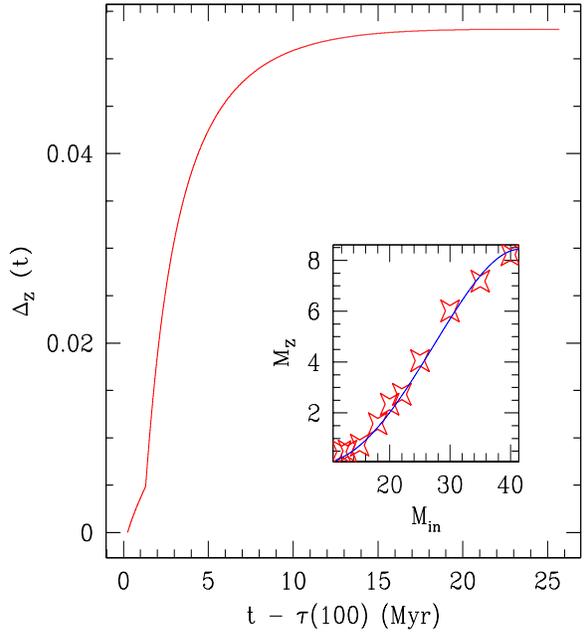, height=10.5cm,width=9cm}
\vspace{-0.5cm}
\caption[]{\label{fig:fig 6} Normalized metal production $\Delta_Z$ 
as a function of time.  $\tau (100)$ is the lifetime of a 100 M$_\odot$ 
star.  In the inner plot is shown the mass $m_Z$ of heavy elements 
ejected by a SN of initial mass $M_{in}$ and our approximation to this 
trend.}
\end{figure}

\begin{table}
\begin{centering}
\caption[]{Normalized metal production $\Delta_Z$ as a function 
of time.}
\begin{tabular}{cc}
\noalign{\smallskip}
\hline
\noalign{\smallskip}
$t - \tau (100)$ (Myr) & $\Delta_Z (t)$ * 100.\\
\noalign{\smallskip}
\hline\noalign{\smallskip}
  1. & 0.367\\
  2. & 1.875\\
  3. & 3.089\\
  4. & 3.797\\
  5. & 4.245\\
  6. & 4.544\\
  7. & 4.572\\
  8. & 4.900\\
  9. & 5.008\\
  10. & 5.087\\
  11. & 5.147\\
  12. & 5.191\\
  13. & 5.224\\
  14. & 5.249\\
  15. & 5.267\\
  16. & 5.281\\
  17. & 5.291\\
  18. & 5.298\\
  19. & 5.303\\
  20. & 5.307\\
  21. & 5.309\\
  22. & 5.309\\
  23. & 5.309\\
  24. & 5.309\\
  25. & 5.309\\
\noalign{\smallskip}
\hline
\end{tabular}
\vspace{0.3cm}

Normalized metal production as a function of time (see eq. 21 for the 
definition of $\Delta_Z (t)$).
\end{centering}
\end{table} 

\subsection{Triggered star formation in the shell}

The expansion of a supershell can trigger star formation in the shell
itself if the onset of gravitational instabilities in the fragments
occurs when the shell is still bound to the PGC potential.  Moreover,
we must require that the time-scale for the onset of these
gravitational instabilities is shorter than the time-scale needed to
reach the tidal radius of the PGC.  In order to calculate the
time-scale needed for the onset of gravitational instabilities, we
follow an approach similar to the one of McCray \& Kafatos (1987),
namely we consider a small circular disk with radius $\xi << r$ on the
surface of the expanding spherical shell, which expands in a $r^{-2}$
density profile with a constant velocity $v_b$.  We indicate with
$\theta \simeq \xi / r$ the half-angle subtended by the disk.  The
thermal, kinetic and gravitational energy of the disk are $m c_s^2$,
${1 \over 2} m \bigl({d \xi \over dt}\bigr)^2$, $- {2 \over 3} G {m^2
  \over \xi}$, respectively.  The mass of the disk is $m(r,\theta) =
\pi \theta^2 \rho_0 r_c^2 (r - r_c)$, thus we obtain: $E_K = {1 \over
  2} \pi \theta^4 v_b^2 \rho_0 r_c^2 (r - r_c)$; $E_{Th} = \pi
\theta^2 \rho_0 r_c^2 (r - r_c) c_s^2$; $E_B = - {2 \over 3} G \pi^2
\theta^3 \rho_0^2 r_c^4 (r - r_c)^2 / r$.  The criterion for the onset
of a gravitational instability is approximatively $E_K + E_{Th} + E_B
< 0$, which transforms into:

\begin{equation}
{1 \over 2} \theta^2 v_b^2 - {2 \over 3} G \pi \theta \rho_0 r_c^2 
{{r - r_c} \over r} + c_s^2 < 0.
\end{equation}
\noindent
The $\Delta$ of this equation must be $>0$, otherwise the above
expression is positive for any value of $\theta$.  This transforms
into the following condition:

\begin{equation}
r > r_c \Biggl\lbrack 1 - {3 c_s v_b \over \sqrt 2 G \pi \rho_0 r_c^2} 
\Biggr\rbrack^{-1},
\end{equation}
provided that $3 c_s v_b < \sqrt 2 G \pi \rho_0 r_c^2$.

This radius is reached at a time 

\begin{equation}
t_{*, 6} = {2.55 \bigl({M_6 \over P_4}\bigr)^{1/6} \eta^{-1/4} 
g(m_l)^{1/2} \over 1 - 11.07 (M_6 P_4)^{-1/2} \eta^{1/4} g(m_l)^{-1/2}}.
\end{equation}
\noindent
The metallicity of this self-enriched population of stars is simply
given by:

\begin{equation}
Z = \gamma M_Z (t_*) / M_{sw} (t_*), 
\end{equation}
\noindent
where $\gamma$ is the {\it mixing efficiency}, namely the fraction of
metals produced in the first generation of stars, able to mix with the
surrounding cold shell in a time-scale shorter than the time-scale
needed for the onset of gravitational instabilities in the shell.

\section{Results}

\subsection{Constraints on the mass of the PGC}

The time needed for the onset of a gravitational instability able to
generate the population of stars we observe nowadays in GCs ($t_{*,
6}$) is given by eq. (26).  This time-scale has to be shorter than the
time-scale needed to reach the tidal radius ($t_{tid, 6}$, given by
eq. (20)), otherwise the energy released by the first generation of
stars simply blows away all the gas in the PGC.  By comparing these
two time-scales, we obtain a condition for the mass of the PGC:

\begin{equation}
M_{Min, 1} > 12.46 P_4^{-1/2} \eta^{1/4} g(m_l)^{-1/2}.
\end{equation}
\noindent
We can obtain another constraint on $M_6$ by considering that, in the
moment in which stars form in the shell, the shell has to be bound.
It means that the potential energy of the shell should be larger than
its kinetic energy, i.e. ${1 \over 2} M_{sw} (t_*) v_b^2 < {1 \over 2}
{{G M_{sw} (t_*)^2} \over R_s (t_*)}$.  In this way we can define a
minimum shell mass:

\begin{equation}
M_{sh \, min, 6} = {9.11 \cdot 10^{-2} \bigl({M_6 \over P_4}\bigr)^{2/3} 
\eta^{1/2} g(m_l)^{-1} \over 1 - 11.07 (M_6 P_4)^{-1/2} \eta^{1/4} 
g(m_l)^{-1/2}}.
\end{equation}
\noindent
The swept-up mass calculated by means of eq. (21) has to be larger
than $M_{sh \, min}$ in order that the shell be bound.  This
translates into another condition for the mass of the PGC, namely:

\begin{equation}
M_{Min, 2} > 0.6 P_4^{-2} \eta^{3/2} g(m_l)^{-3}.
\end{equation}
\noindent
As usual, all the masses (eqs. 28, 29 and 30) are in units of 10$^6$
M$_\odot$.

\begin{figure}
\centering
\epsfig{file=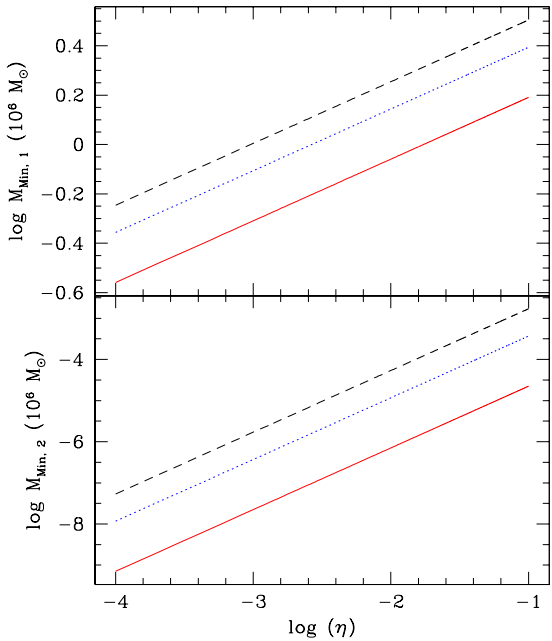, height=9cm,width=9cm}
\caption[]{\label{fig:fig 7} Lower limits for the mass of the PGC
  calculated according to eq. (28) (upper panel) and eq. (30) (lower
  panel).  The thermalization efficiency $\eta$ varies with
  continuity, whereas the lower IMF mass $m_l$ is equal to 0.1 (solid
  line), 1 (dotted line), 3 (dashed line).}
\end{figure}

\begin{figure}
\centering
\epsfig{file=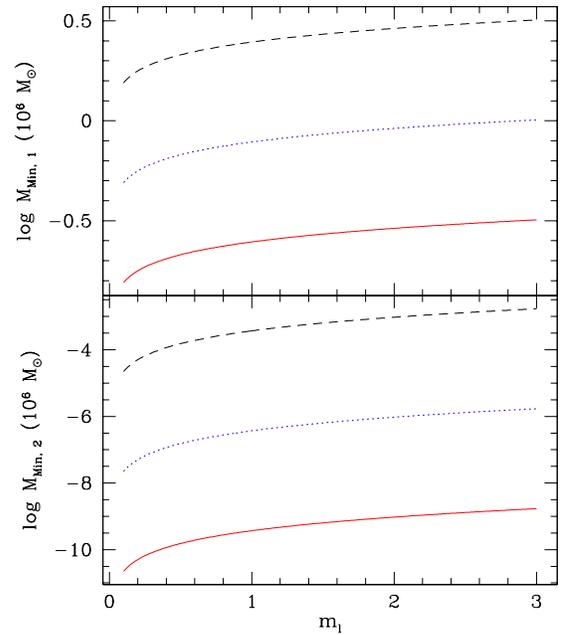, height=9cm,width=9cm}
\caption[]{\label{fig:fig 8} As Fig. 6 but with varying lower IMF mass
$m_l$.  The thermalization efficiency $\eta$ is equal to 10$^{-5}$
(solid line), 10$^{-3}$ (dotted line), 0.1 (dashed line).}
\end{figure}

The resulting dependence of the minimum mass of the PGC calculated
according to eq. (28) ($M_{Min 1}$) and eq. (30) ($M_{Min, 2}$) is
shown in Fig. 7 (as a function of $\eta$) and in Fig. 8 (as a function
of $m_l$).  For simplicity, the external pressure is kept constant in
these plots and in the following sections.  It is assumed to be $P_4 =
10$, in agreement with the pressure of the hot protogalactic
background (Fall \& Rees 1985; Murray \& Lin 1992).  We analyze the
dependence of [Fe/H] on $P_4$ in Sect. 4.2.4.  These plots show that
for any reasonable choice of parameters, the critical mass required in
order to get the shell bound to the PGC potential well ($M_{Min, 2}$)
is always extremely low.  This is therefore not a severe constraint on
the minimum mass of the PGC.

The constraint based on the time needed for the onset of a
gravitational instability compared to the time required for the shock
to reach the tidal radius (expressed by eq. (28)) is more stringent
and led us to consider only PGC with masses larger than $\sim$ 10$^6$
M$_\odot$.  This is the same threshold mass found by SW00 and is also
comparable with the threshold mass found by Brown, Burkert \& Truran
(1995), but is significantly larger than the one found by Morgan \&
Lake (1989).

\subsection{The Magnitude-Metallicity relation in GCs}

A direct comparison between the results of this simplified model and
the observations can be made as follows.  We can assume the same
efficiency of star formation adopted in Sect. 3.2, namely that 30\% of
the mass of the shell at the moment of the onset of the gravitational
instability ($M_{sw} (t_*)$) transforms into stars.  This is therefore
the mass of our newly formed GC.  This should be considered as a lower
limit, since, after the formation of the second, polluted generation
of stars, the GC is still rich in gas.  Shocks created by the
supernovae exploding in the shell can still compress the gas, leading
to the formation of new stars.

To convert the mass in stars into visual magnitude, we assume a
constant mass-to-light ratio M/L$_V$ $=1.5$, typical for GC systems in
the Milky Way (Harris 1996), M31 (Djorgovski et al. 1997) and M33
(Larsen et al. 2002).  At this point the visual magnitude is simply
calculated by means of the following formula:

\begin{equation}
M_V = 5.41 - 2.5 \cdot \log \bigl\lbrack{0.2 \cdot M_{sw} (t_*)}
\bigr\rbrack.
\end{equation}
\noindent
Now we can directly compare the observed magnitude-metallicity
relation with the results of our approximations.  We take the sample
of data already shown in Fig. 3, namely the GCs of the Milky Way with
$|b| > 20$, in order to avoid the effect of the mass loss due to
stripping passing through the disk of the Galaxy.  We explore the
parameter space, varying the mixing efficiency $\gamma$, the
thermalization efficiency $\eta$, the lower limit of the IMF
distribution $m_l$ and the external pressure $P_4$.

\subsubsection{The Magnitude-Metallicity relation as a function of
$\gamma$}

The mixing between the metals produced in the first generation of
stars and the swept-up shell is a complex process, involving poorly
known physics.  In the idealized wind-blown bubble model, the shocked
wind (hot, central region filled with metals) and the shocked ISM (the
shell of unpolluted swept-up material) are separated by a contact
discontinuity, thus in principle no flux of matter is possible between
these two regions.  Actually many physical processes can allow a
mixing (mixing layers, turbulence, thermal instabilities, thermal
conduction), but these processes are poorly constrained.  We therefore
decided to keep the mixing efficiency $\gamma$ as a free parameter,
trying to find some constraint on its value through the comparison of
our results with the observations.

\begin{figure}
\centering
\vspace{-1cm}
\epsfig{file=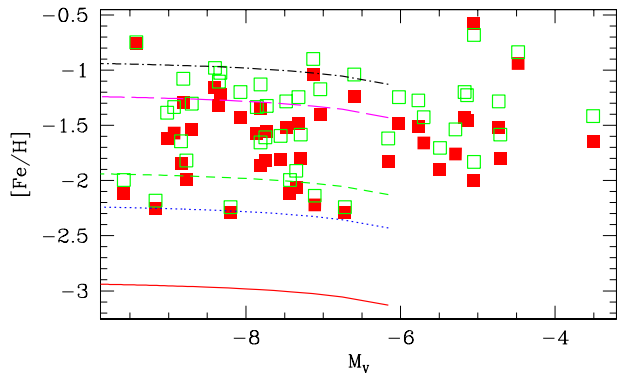, height=9cm,width=9cm}
\vspace{-1cm}
\caption[]{\label{fig:fig 9} Observed Magnitude-Metallicity relation
for a selected sample of GCs (see Fig. 3) compared with model results.
The mixing efficiency $\gamma$ is taken as 0.01 (solid line), 0.05
(dotted line), 0.1 (dashed line), 0.5 (long-dashed line) and 1
(dot-dashed line).}
\end{figure}

The Magnitude-Metallicity relation as a function of $\gamma$ is shown
in Fig. 9.  In this set of models, the thermalization efficiency
$\eta$ and the lower limit of the IMF distribution $m_l$ are set to be
10$^{-4}$ and 1, respectively.  Although the value of the
thermalization efficiency is rather low, it is fully consistent with
the value found in eq. (10).  As we can see in this figure, the
minimum required mixing efficiency is between 0.05 and 0.1, whereas,
in order to reach the most metal-rich clusters, a very large $\gamma$
is required.  A non-negligible mixing between the hot cavity, filled
with metals, and the shell, should therefore occur, but the magnitude
of this process can be only poorly constrained due to the huge spread
in the observed metallicities of GCs.  Owing to the variety of
processes contributing to the mixing of metals with the shell, the
spread in metallicity can reflect, at least in part, a
cluster-to-cluster variation of $\gamma$.

%For example, Recchi et al. (2002) found a
%mixing efficiency depending on the adopted parameters and on the
%adopted Star Formation History taken into consideration.

It is also worth noting that the evolution of the metallicity [Fe/H]
as a function of M$_V$ is almost flat.  This is in agreement with the
fact that no evident trend is shown in the observed M$_V$ vs. [Fe/H]
relationship.

A similar comparison of the results of the models with the
observations can be made taking into consideration the masses of GCs
instead of the magnitudes.  In this case, we do not have to use eq.
(31) but we can directly compare the swept-up mass at the moment of
the onset of the gravitational instabilities with the mass tabulated
by Pryor \& Meylan (1993) (see Fig. 4).  The results are plotted in
Fig. 10.  The conclusions we can draw from this plot are the same: a
very low mixing efficiency can be ruled out and, in order to reproduce
the GCs with the largest metallicities, we have to assume that more
than 50 \% of the metals mix with the surrounding shell in a
time-scale shorter than the time needed for the onset of gravitational
instabilities.

\begin{figure}
\centering
\vspace{-1cm}
\epsfig{file=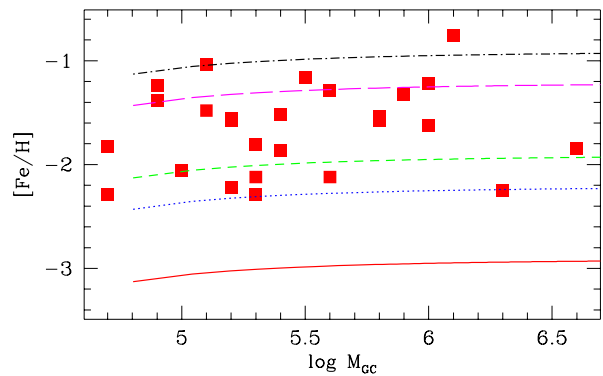, height=9cm,width=9cm}
\vspace{-1cm}
\caption[]{\label{fig:fig 10} Observed Mass-Metallicity relation for a
selected sample of GCs (see Fig. 4) compared with model results.  The
mixing efficiency $\gamma$ is taken as 0.01 (solid line), 0.05 (dotted
line), 0.1 (dashed line), 0.5 (long-dashed line) and 1 (dot-dashed
line).}
\end{figure}

\subsubsection{The Magnitude-Metallicity relation as a function of
$m_l$}

Many authors believe that the IMF of primordial objects should have
been top-heavy, owing to the fact that the CMB radiation keeps the gas
warm enough to raise the Jeans mass.  We therefore varied the lower
limit of the IMF distribution $m_l$ in order to see what effect an IMF
biased towards massive stars can have in the metallicity of GCs.  

\begin{figure}
\centering
\vspace{-1cm}
\epsfig{file=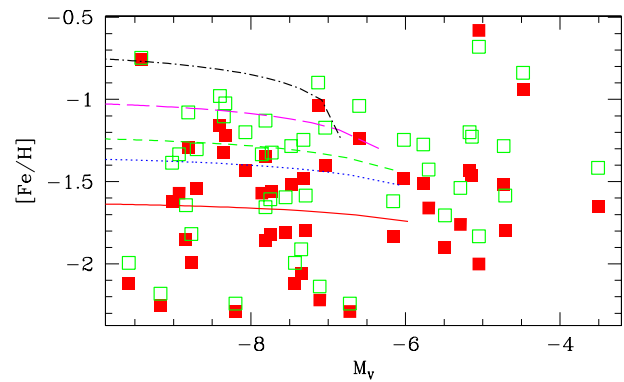, height=9cm,width=9cm}
\vspace{-1cm}
\caption[]{\label{fig:fig 11} Same as Fig. 9 but with varying lower
mass $m_l$.  This parameter ranges the values 0.1 (solid line), 0.5
(dotted line), 1 (dashed line), 3 (long-dashed line) and 10
(dot-dashed line).}
\end{figure}

The Magnitude-Metallicity relation as a function of $m_l$ is shown in
Fig. 11.  In this set of models, the thermalization efficiency $\eta$
and the mixing efficiency $\gamma$ are set to be 10$^{-4}$ and 0.5,
respectively.  As we can see in this figure, an IMF strongly biased
towards massive stars ($m_l = 10$) can justify the most metal-rich GC
observed.  If the IMF is so strongly top-heavy, mixing efficiencies
larger that 0.5 cannot be allowed.  This mixing efficiency is
consistent with the value found by Recchi et al. (2001) in their
simulations of IZw18.  Owing to the fact that the selected sample of
GCs is not perfectly coeval, small cluster-to-cluster variations of
$m_l$ are possible, explaining therefore, at least in part, the
observed scatter.

Also for these models, [Fe/H] does not vary so much with $M_V$.  Only
for models with large $m_l$ a correlation arises, in the sense that
the faintest (and less massive) GCs are the most metal-rich.  This is
due to the fact that $t_*$ grows with $M_6$.  At large $t$, $M_Z$ is
almost independent on $t$ (see Fig. 6), whereas $M_{sw}$ grows
linearly with $t$ (eq. 21).  From eq. (27) we can see that Z should
decrease.  This trend has been found also by P99.  Statistical
properties of a selected sample of old GCs (the so-called Old Halo
GCs) seem to confirm this trend (Parmentier \& Gilmore 2001), although
the scatter is huge.

\subsubsection{The Magnitude-Metallicity relation as a function of
$\eta$}

The thermalization efficiency $\eta$ is the ratio of the mechanical
energy of a single SN to the internal energy gained by the ISM after
the explosion.  As we have seen in Sect. 3.2, the value of $\eta$ can
be estimated analytically (eq. 9) but, due to the large uncertainties
in this evaluation, we decided to keep the thermalization efficiency
as a free parameter.  However, owing to the very large densities in the
core of PGCs, this value should be very low.  We therefore decided to
span the parameter range [5 $\cdot$ 10$^{-3}$, 10$^{-6}$]. 

\begin{figure}
\centering
\vspace{-1cm}
\epsfig{file=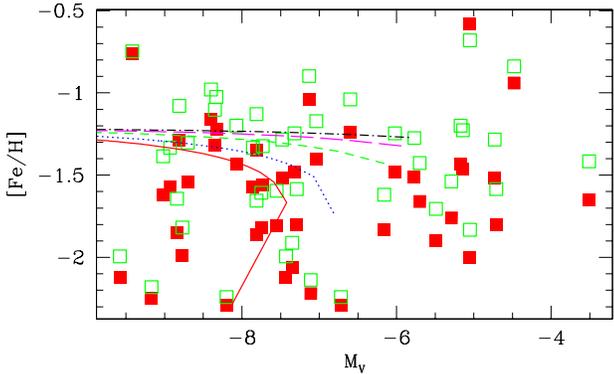, height=9cm,width=9cm}
\vspace{-1cm}
\caption[]{\label{fig:fig 12} Same as Fig. 9 but with varying
thermalization efficiency $\eta$.  This parameter ranges the values 5
$\cdot$ 10$^{-3}$ (solid line), 10$^{-3}$ (dotted line), 10$^{-4}$
(dashed line), 10$^{-5}$ (long-dashed line) and 10$^{-6}$ (dot-dashed
line).}
\end{figure}

The Magnitude-Metallicity relation as a function of $\eta$ is shown in
Fig. 12.  In this set of models, the lower limit of the IMF
distribution $m_l$ and the mixing efficiency $\gamma$ are set to be 1
and 0.5, respectively.  As we can see in this figure, below $\eta =
10^{-4}$, the dependence of the metallicity with $\eta$ is very weak.
Above this value, the dependence becomes stronger due to the
non-linear dependence of $t_*$ (and consequently $M_Z (t_*)$ and
$M_{sw} (t_*)$) on $\eta$ (see eq. 26).

\subsubsection{The Magnitude-Metallicity relation as a function of
$P_4$}

So far we have analyzed the dependence of the M$_V$ vs. [Fe/H]
relation as a function of internal parameters of the PGC.  Of course
also external parameters play a role. In this subsection we analyze
the evolution of the metallicity of GCs as a function of the external
pressure $P_4$.  This dependence is shown in Fig. 13.

\begin{figure}
\centering
\vspace{-1cm}
\epsfig{file=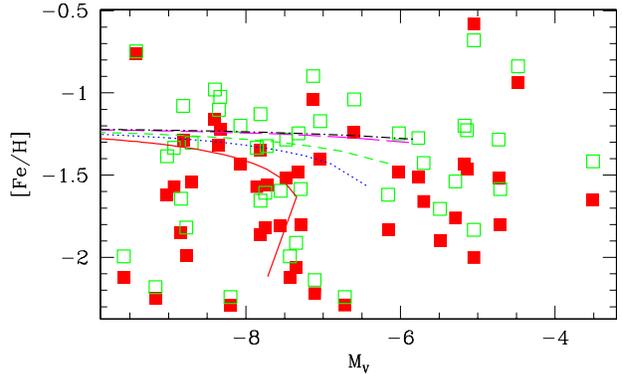, height=9cm,width=9cm}
\vspace{-1cm}
\caption[]{\label{fig:fig 13} Same as Fig. 9 but with varying external
pressure $P_4$.  This parameter ranges the values 1 (solid line), 5
(dotted line), 10 (dashed line), 50 (long-dashed line) and 100
(dot-dashed line).}
\end{figure}

The effect of the external pressure is similar to the effect of the
thermalization efficiency $\eta$, in the sense that above $P_4$ $=$ 10
this effect is negligible and the tracks lie close to each other.  For
low values of the external pressure, strongly non-linear effects on
the eq. (26) create the knee observed for the model with $P_4 = 1$
(solid line) and $P_4 = 5$ (dotted line).  Smaller values of
$P_4$ violate the condition $3 c_s v_b < \sqrt 2 G \pi \rho_0 r_c^2$
(see Sect. 3.5) and cannot be taken into account in this simplified
approach.  As stated in Sect. 4.1, these low pressure environments are
unlikely in protogalactic backgrounds.

\section{Conclusions}

We performed an analytical study of the evolution of a proto-globular
cluster (PGC) assuming an initial burst of star formation occurring
inside the core radius of the initial gaseous distribution.  We
followed the evolution of the shock wave formed after energy release
from the first SNeII and we considered the possibility that
gravitational instabilities may arise in the swept-up shell, leading
to the formation of a second population of stars, with metallicity
larger than zero, owing to the pollution from the ejecta of the first
supernovae.  Our results can be summarized as follows:

\begin{itemize}

\item Is it possible to calculate a lower mass of the PGC assuming
that the onset of the gravitational instability able to produce the
second generation of stars should occur before the shell reaches the
edge of the PGC gaseous distribution.  With reasonable choices of the
parameters, this threshold mass is of the order of 10$^6$ M$_\odot$,
in agreement with other similar studies (SW00, Brown et al. 1995).

\item In order to produce a second stellar population with a
metallicity in agreement with the one observed in the most metal-poor
GCs, it is necessary to mix at least 5 -- 10 \% of the metals produced
in the central burst of star formation with the surrounding swept-up
shell.  The most metal-rich GCs are consistent with a very large
mixing efficiency (larger than 0.5; i.e. more than 50\% of the metals
mix with the surrounding shell).

\item The evolution of theoretical tracks in the M$_V$ vs. [Fe/H] plot
is almost constant.  This is consistent with the fact that no obvious
trend can be found in the observed M$_V$ vs. [Fe/H] relationship.  The
scatter of the metallicity is therefore due mostly to scatter in the
internal parameters of the cluster.

\item A top-heavy IMF (with lower mass $m_l$ larger than 1 M$_\odot$)
seems to be required in order to explain the GCs with the largest
metallicities.  

\item The huge spread in the metallicity observed in the sample of GCs
can be explained, at least in part, with cluster-to-cluster variations
of the structural parameters $\gamma$, $\eta$ and $m_l$, as well as
variations of the external pressure $P_4$.

\end{itemize}

\begin{acknowledgements}
  
  We acknowledge an anonymous referee for suggestions and comments
  which improved the paper.  S.R. acknowledges generous financial
  support from the Alexander von Humboldt Foundation and Deutsche
  Forschungsgemeinschaft (DFG) under grant HE 1487/28-1 and from the
  INAF (Italian National Institute for Astrophysics) - Osservatorio
  Astronomico di Trieste

\end{acknowledgements}

\end{document}